\documentclass[trackchanges,resetfootnote]{aastex701}

\newcommand{\gaia}{\it Gaia}
\newcommand{\kepler}{\it Kepler}
\defcitealias{gaiadr3}{Gaia Collaboration et al. 2025}
\begin{document}

\title{Mining the {\kepler} Field: Atmospheric Parameters, Bolometric Corrections, and Luminosities}

\author[orcid=0000-0003-4556-1277,gname=Diego,sname=Godoy-Rivera]{Diego Godoy-Rivera}
\altaffiliation{Juan de la Cierva Fellow}
\affiliation{Instituto de Astrof\'isica de Canarias (IAC), C/V\'ia Lactea, s/n, E-38205 La Laguna, Tenerife, Spain}
\affiliation{Universidad de La Laguna (ULL), Departamento de Astrof\'isica, E-38206 La Laguna, Tenerife, Spain}
\email[show]{diego.godoy.rivera@iac.es}  

\author[orcid=0000-0001-6529-9769,gname=Desmond,sname=Grossmann]{Desmond H. Grossmann}
\altaffiliation{"la Caixa" INPhINIT Doctoral Fellow}
\affiliation{Instituto de Astrof\'isica de Canarias (IAC), C/V\'ia Lactea, s/n, E-38205 La Laguna, Tenerife, Spain}
\affiliation{Universidad de La Laguna (ULL), Departamento de Astrof\'isica, E-38206 La Laguna, Tenerife, Spain}
\email{desmond.grossmann@iac.es}

\author[orcid=0000-0003-1290-3621,gname=Tyler, sname=Richey-Yowell]{Tyler Richey-Yowell} 
\altaffiliation{Percival Lowell Postdoctoral Fellow}
\affiliation{Lowell Observatory, Flagstaff, AZ 86004, USA}
\email{try@lowell.edu}

\author[orcid=0000-0001-7195-6542,gname=Ângela,sname=Santos]{Ângela R. G. Santos}
\affiliation{Instituto de Astrofísica e Ciências do Espaço, Universidade do Porto, CAUP, Rua das Estrelas, PT4150-762 Porto, Portugal}
\affiliation{Departamento de Física e Astronomia, Universidade do Porto, Rua do Campo Alegre 687, PT4169-007 Porto, Portugal}
\affiliation{Université Paris-Saclay, Université Paris Cité, CEA, CNRS, AIM, 91191, Gif-sur-Yvette, France}
\email{angela.santos@cea.fr}

\author[orcid=0000-0002-0129-0316,gname=Savita,sname=Mathur]{Savita Mathur}
\affiliation{Instituto de Astrof\'isica de Canarias (IAC), C/V\'ia Lactea, s/n, E-38205 La Laguna, Tenerife, Spain}
\affiliation{Universidad de La Laguna (ULL), Departamento de Astrof\'isica, E-38206 La Laguna, Tenerife, Spain}
\email{smathur@iac.es}

\author[orcid=0000-0002-8854-3776,gname=Rafael,sname=García]{Rafael A. García}
\affiliation{Université Paris-Saclay, Université Paris Cité, CEA, CNRS, AIM, 91191, Gif-sur-Yvette, France}
\email{rafael.garcia@cea.fr}

\begin{abstract}
The $\sim$ 200,000 stars observed by the {\kepler} mission have provided unprecedented constraints across astrophysics. With the advent of modern spectroscopic and photometric surveys, new limits in stellar characterizations are within reach. In this work, we report a compilation of atmospheric parameters ($T_{\text{eff}}$, $\log(g)$, and [M/H]) for the {\kepler} stars by crossmatching with several spectroscopic and spectro-photometric surveys. We use these to calculate bolometric corrections, which combined with color-magnitude diagram (CMD) information from {\gaia} yield self-consistent luminosities on a survey-by-survey basis. These properties will aid in future explorations of {\kepler} data towards new astrophysical insights. We make our catalog publicly available online in Zenodo (doi:10.5281/zenodo.18620911).
\end{abstract}

\phantomsection

\section{Introduction}
\label{sec:introduction}

The light curves from {\kepler} \citep{borucki10} have enabled discoveries in the fields of exoplanets, asteroseismology, stellar rotation and activity, among others. To fully exploit the {\kepler} data, thorough stellar characterizations are needed. While our knowledge of its targets has increased in the last decade, the advent of modern surveys allows us to push the limits in stellar parameter determinations.

We performed a crossmatch of the $\sim$ 200,000 {\kepler} stars with modern surveys. We gathered atmospheric parameters, computed bolometric corrections, and combined them with {\gaia} DR3 data \citepalias{gaiadr3} to calculate luminosities, a crucial astrophysical property.

\section{Method}
\label{sec:method}

We defined the target sample as the 196,762 {\kepler} stars with {\gaia} DR3 counterparts from \citet{godoyrivera25}. We calculated luminosities and uncertainties as
\begin{equation}
\left(\frac{L}{\text{L}_{\odot}}\right) = 10^{\left(\frac{M_{G_0} + \text{BC}_{G} - \text{M}_{\text{bol},\odot}}{-2.5}\right)},
\label{eqn:Lbol_central}
\end{equation}
and
\begin{equation}
\left(\frac{\sigma_L}{L}\right) =  \left(\frac{\ln(10)}{2.5}\right)\sqrt{\sigma^2_{M_{G_0}}+\sigma^2_{\text{BC}_{G}}},
\label{eqn:Lbol_sigma}
\end{equation}
where $M_{G_0}$ is the de-reddened $G$-band absolute magnitude, $\text{BC}_{G}$ is the $G$-band bolometric correction, and $\text{M}_{\text{bol},\odot} =4.74$ mag is the solar bolometric magnitude \citep{mamajek15}. 

We took the absolute magnitudes from \citet{godoyrivera25}, who used the {\gaia} DR3 parallaxes and extinction-corrected photometry, to place the targets on the absolute and de-reddened color-magnitude diagram (CMD). After quality cuts, their CMD sample comprised 179,295 stars, with a median error of $\sigma_{M_{G_0}}=0.063$ mag. Bolometric corrections were calculated following \citet{creevey23} with the \texttt{gaiadr3\_bcg}\footnote{\url{https://gitlab.oca.eu/ordenovic/gaiadr3_bcg}} function, which takes as input the atmospheric parameters ($T_{\text{eff}}$, $\log(g)$, [Fe/H], [$\alpha$/Fe]) and calculates $\text{BC}_G$ by interpolating MARCS synthetic spectra \citep{gustafsson08}.

\section{Spectroscopic Compilation}
\label{sec:compilation_spectroscopic}

\begin{figure*}
\centering
\includegraphics[width=17.0cm]{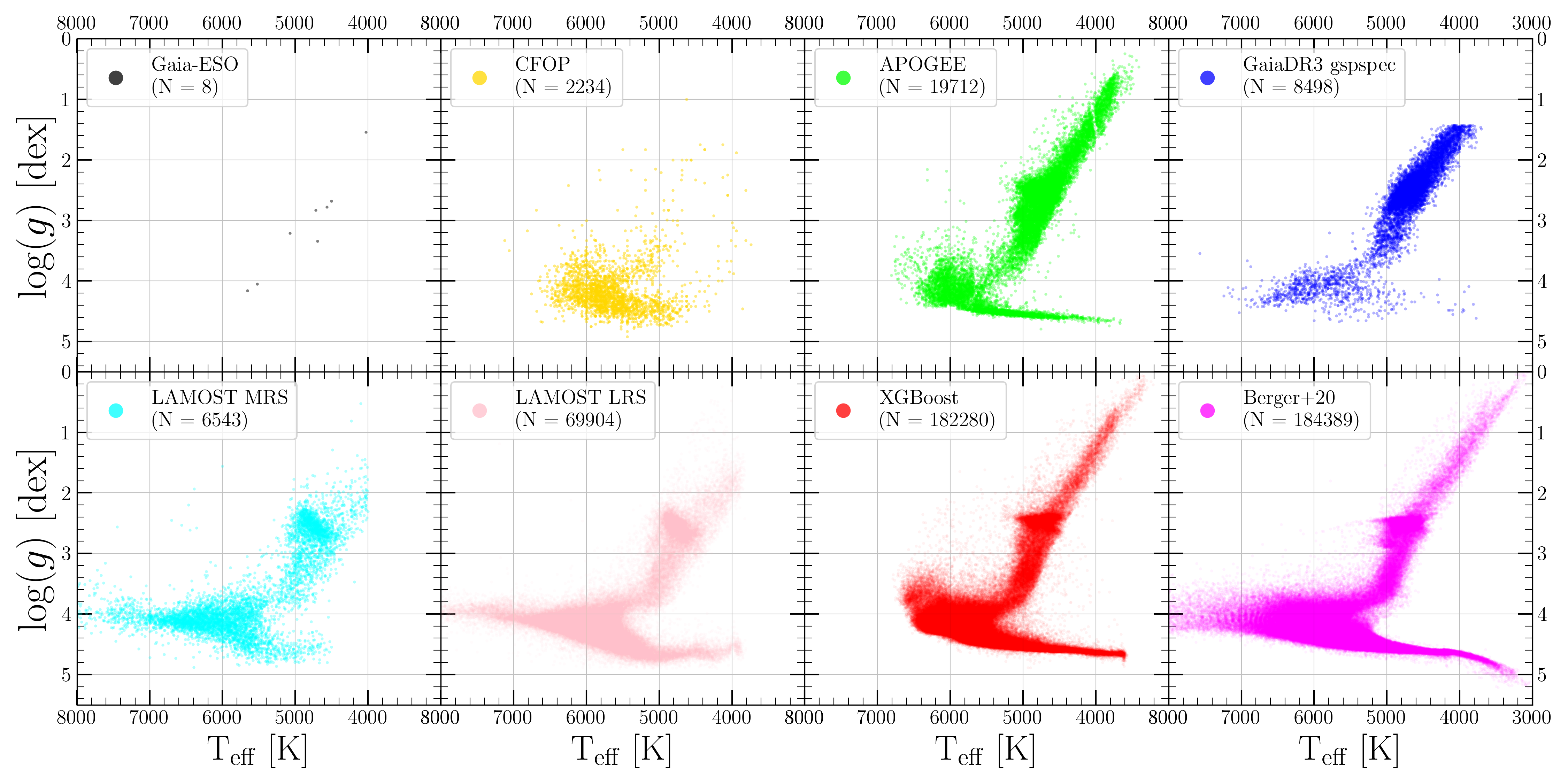}\\
\includegraphics[width=17.4cm]{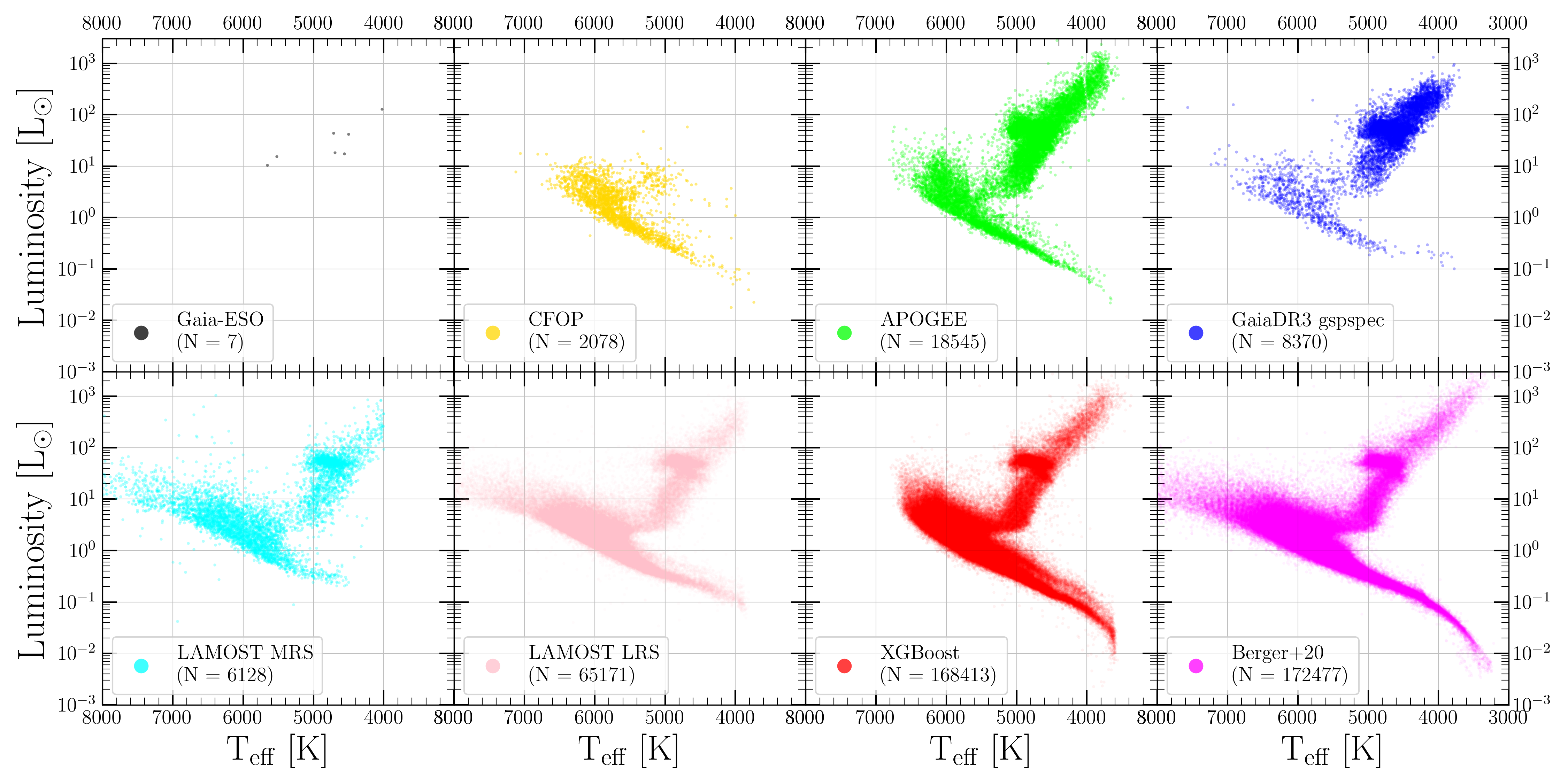}\\
\caption{Kiel (top) and Hertzsprung-Russell (bottom) diagram for the {\kepler} stars. Each panel indicates a survey, with the number of stars found in each indicated in the legend.}
\label{fig:Kiel_HRD}
\end{figure*}

Given the need for atmospheric parameters in the computation of bolometric corrections, we compiled these from surveys. The crossmatch is available online and summarized in the Kiel diagram of Fig.~\ref{fig:Kiel_HRD}.

Sorting by resolution, we crossmatched with: {\gaia}-ESO (v5.1\footnote{All crossmatches found are members of NGC~6791 \citep{hunt23}.}, $R \sim$ 47,000; \citealt{hourihane23}), the {\kepler} Community Follow-up Observation Program (CFOP, $R \sim$ 40,000; \citealt{furlan18}), APOGEE (DR17, $R \sim$ 22,000; \citealt{abdurrouf22}), {\gaia} DR3 \texttt{gspspec} ($R \sim$ 11,500; \citealt{recioblanco23}), and LAMOST (DR10 v2.0; \citealt{cui12}) Medium- and Low-Resolution Spectroscopy (MRS, $R \sim$ 7,500; LRS, $R \sim$ 1,800). We also queried spectro-photometric catalogs, namely XGBoost ($R \sim$ 100, \citealt{andrae23}), which was based on the {\gaia} DR3 XP coefficients and trained on the APOGEE data, and \citet{berger20}, who reported stellar parameters for the {\kepler} stars based on isochrones, broadband photometry, and {\gaia} DR2 parallaxes.

For reliability we applied quality cuts. For CFOP, we removed stars with Kiel diagram locations unpopulated by other surveys or unphysical values (KIC 8823868, KIC 9552608, and KIC 2437209). For APOGEE, we removed targets with the \texttt{ASPCAPFLAG} \texttt{STAR\_WARN}, \texttt{STAR\_BAD}, or \texttt{M\_H\_WARN} quality flags. For {\gaia} DR3 \texttt{gspspec}, we took the calibrated best-quality sample from \citet{godoyrivera25}. For LAMOST, we only considered targets with measured ($T_{\text{eff}}$, $\log(g)$, [M/H]) parameters and uncertainties. For XGBoost, we did not consider the targets with {\gaia} color $(BP-RP) \leq 0.5$ mag, or with $(BP-RP) \geq 2.5$ mag and $\log(g) > 3.6$ dex, as in those regimes their training set had limited coverage and their data did not follow the expected color-temperature relation. For \citet{berger20}, we only kept targets with goodness-of-fit $>$ 0.99. These quality cuts did not significantly impact the amount of successful crossmatches.

We adopted the uncertainties reported in the respective surveys. For {\gaia} DR3 \texttt{gspspec}, we averaged the (largely symmetric) lower and upper error bars to streamline the uncertainty propagation. For XGBoost, no star-by-star uncertainties were reported, and we adopted their overall mean precisions as uncertainties (50 K in $T_{\text{eff}}$, 0.08 dex in $\log(g)$, and 0.1 dex in [M/H]).

As shown in Fig.~\ref{fig:Kiel_HRD}, the fraction of the {\kepler} stars found varies strongly on a survey-by-survey basis. We performed inter-survey comparisons, finding varying degrees of agreement (e.g., median $\Delta T_{\text{eff,(LAMOSTLRS-APOGEE)}} \approx +6$ K  and $\Delta T_{\text{eff,(GaiaDR3gspspec-APOGEE)}} \approx -34$ K). While preferable, homogenizing parameters across surveys is non-trivial \citep{thomas24} and beyond the scope of this work. 

\section{Bolometric Corrections}
\label{sec:bolometric_corrections}

We calculated bolometric corrections using the compiled atmospheric parameters\footnote{$\sim$ 100 stars with atmospheric parameters fall outside the input range of the \texttt{gaiadr3\_bcg} function and lack bolometric corrections.}. While some surveys reported [Fe/H] and others [M/H], we took both of them to represent metallicity. We assumed [$\alpha$/Fe]=0 as this parameter is scarcely available and tests showed it has little impact on the resulting $\text{BC}_{G}$ values. For each star we performed 1000 Monte Carlo simulations of the atmospheric parameters assuming Gaussian distributions (R. A. García et al. in preparation), and adopted the median and its $1\sigma$ range of the resulting distribution as $\text{BC}_G$ and $\sigma_{\text{BC}_G}$. The median error is  $\sigma_{\text{BC}_G} = 0.010$ mag, although with significant survey-dependent trends.

\section{Luminosities}
\label{sec:bolometric_luminosities}

For stars with available absolute magnitudes and bolometric corrections, we used Equations (\ref{eqn:Lbol_central}) and (\ref{eqn:Lbol_sigma}) to calculate luminosities. The median luminosity uncertainty is $\left(\frac{\sigma_L}{L}\right)=5.7\%$. The resulting Hertzsprung-Russell diagram is shown in Fig.~\ref{fig:Kiel_HRD}. The sample includes the upper main-sequence (MS), solar-like stars, and evolved phases (e.g., red giant branch and red clump). Heterogeneities can be seen due to the inherited atmospheric parameters from different surveys. This is particularly evident along the MS for XGBoost, and we thus caution about its use in this regime. Our luminosities are in good agreement with the literature (e.g., 93\% of the LAMOST LRS values are within 3$\sigma$ of {\gaia} DR3 FLAME). We make our catalog publicly available as a resource to the community.

\section*{Data Availability}
\label{sec:data_availability}
The catalog is available in electronic form on Zenodo at \href{https://doi.org/10.5281/zenodo.18620911}{doi:10.5281/zenodo.18620911} and reports the atmospheric parameters, bolometric corrections, and luminosities.

\begin{acknowledgments}
We thank Santi Cassisi, Alejandro Martín Escabia, Travis Berger, and Thomas Masseron for helpful discussions.
\end{acknowledgments}

\bibliography{kepler_speccomp_bcg_lbol}{}
\bibliographystyle{aasjournalv7}
\end{document}